# Critical trees: counterexamples in model checking of CSM systems using CBS algorithm

by
Wiktor B. Daszczuk

wbd@ii.pw.edu.pl

**Summary**

The important feature of temporal model checking is the generation of counterexamples. In the report, the requirements for generation of counterexample (called critical tree) in model checking of CSM systems are described. The output of TempoRG model checker for QsCTL logic (a version of CTL) is presented. A contradiction between counterexample generation and state space reduction is commented.

# 1. Introduction

The advantage of temporal model checking over other system verification techniques is that the result of any evaluation is guaranteed to be calculated in a finite time [Clar86, Clar89, McMi92]. Yet the result – if it is false – is not sufficient to localize errors in the project. The output of the temporal verifier should support the process of design improvement. The main weapon offered by the temporal verifiers against design errors is the ability to generate counterexample in a case of negative result of evaluation.

Counterexamples are handy for many kinds of reasoning on concurrent systems. The simplest way is the redesign of the system that has been found to be erroneous. Other approach is automatic test generation for verified system [Amma98]. Counterexamples are also used to test the feasibility of errors found during model checking of abstract models of a system [Pasa01].

We will show the manner of finding counterexamples in "classical" evaluation algorithms (finding a fixed point of some functional and automaton-based evaluation). Then the rules of critical tree construction in the CBS algorithm [Dasz01] will be described. An example of the critical tree constructed by the TempoRG program [Dasz02] for example formula will be analyzed. It will be explained why the critical tree should be constructed over full state space rather than on reduced one.

# 2. CRITICAL SEQUENCE AND CRITICAL SUBGRAPH IN "CLASSICAL" EVALUATION ALGORITHMS

In the first group of "classical" methods of evaluation (based on finding a fixed point of some functional [McMi92, Jans98]), a special algorithm [Clar99] is used for finding (path quantifiers are skipped as the rules refer to both LTL and CTL):
- a witness (a state in which a nested formula or formulas hold) when existential modality is used ($F$, $U_s$) and the result is true,
- a counterexample (a state in which a nested formula or formulas do not hold) when general modality is used ($G$, $U_w$, $X$, $X_a$) and the result is false.

In order to provide a better insight into the verified system's behavior, for every individual operator, a sequence of states should be found from the starting state of evaluation to:
- $F \varphi$ - a state in which $\varphi$ holds (witness),
- $\varphi U_s \psi$ - a state in which $\psi$ holds (witness), and $\varphi$ is satisfied from starting state to the predecessor of the witness,
- $G \varphi$ - a state in which $\varphi$ does not hold (counterexample),
- $\varphi U_w \psi$ - a state in which neither $\varphi$ nor $\psi$ holds (counterexample), and $\varphi$ is satisfied from starting state to the predecessor of the counterexample,
- $X \varphi$ - a successor of starting state in which $\varphi$ does not hold (counterexample),
- $X_a \varphi$ - a state having projection onto a successor of starting state in automaton $a$, in which $\varphi$ does not hold (counterexample).

The algorithm is described to find a sequence of states from the starting state to the witness or counterexample [Clar99]. It will be referred to as a **critical sequence**. The disadvantage is that counterexamples are not defined for existential modality. Also, even for general modality if the result is true, then counterexample is not defined (yet the formula may be negated and the positive result may be negative in fact).

Other evaluation methods (tableau for LTL [Jans98, McMi92, Vard96] or alternating automata for CTL and CTL* [Kupf98, Viss97, Viss00, Vard98]) are based on the product of the specific automaton (representing the negation of desired feature) with the state space of the system. If the product is not empty, it is simply the image of all improper behaviors (following the "wrong behavior" automaton). It is the **critical graph** rather than a critical sequence. The advantage of this method is that the critical graph is obtained as a "side effect" of the evaluation [Clar99]. The disadvantage is that the size of the critical graph may be very large and hard to analyze (for example, if **$AG$** $\varphi$ is evaluated and only a few states actually hold $\varphi$, the critical graph contains almost all the state space). Moreover, it is hard to extract the parts of critical tree responsible for given subformulas. The size of the counterexample may be reduced by choosing a single path in the critical graph arbitrarily.

## 3. CRITICAL TREE IN **CBS** ALGORITHM

A special attention was paid to the construction of counterexample in CBS evaluation method [Dasz01]. It cannot be obtained as easy as in tableau method. The second difficulty comes from optimizations: the state in which it is decided that the formula is false need not be exactly the state not satisfying the nested formula.

The assumptions made for finding the critical subgraph are the following:
- only false result of verification gives the critical tree;
- subgraphs responsible for all subformulas influencing the total false result have to be found (regardless of type of modalities and with negations possible);
- only one sequence for every element of the formula (atomic Boolean formula or an operator) is presented to the designer;
- the tree begins in the starting state $s_0$;
- the state finishing the sequence for embracing formula starts the sequence for nested formula (sequences must "stick");
- for two-argument operators (**$Uw$**, =>, * etc.), if both arguments influence the result, sequences for both are presented (for example in a case of $\varphi => \psi$ formula negatively evaluated – it should be explained why the left argument $\varphi$ is true and why $\psi$ is false).

The above requirements assure that the critical subgraph has the form of the **critical tree** [1]. For every operator and every desired result: false or true (opposite to the actual result) the construction of the sequence and rules for the arguments are defined. The rules are collected in the table below.

| No | Formula | Desired result (opposite to the actual result) | State finishing the sequence | Desired result for arguments in the state finishing the sequence |
|----|---------|------------------------------------------------|------------------------------|------------------------------------------------------------------|
| 1 | $\neg\varphi$ | false | starting state $s_0$ | true |

---
[1]  The result may not be a 'regular' tree in fact (a state may occur in more than one sequence). But it is really the tree if we take as elements of the tree the sequences responsible for individual subformulas, indexed by identifiers of these subformulas (see Fig. 2).

| 2 | $\neg\varphi$ | true | starting state $s_0$ | false |
|---|---|---|---|---|
| 3 | $\varphi+\psi$ | false | starting state $s_0$ | $\varphi$ false if it is true <br> $\psi$ false if it is true |
| 4 | $\varphi+\psi$ | true | starting state $s_0$ | $\varphi$ true <br> $\psi$ true |
| 5 | $\varphi*\psi$ | false | starting state $s_0$ | $\varphi$ false <br> $\psi$ false |
| 6 | $\varphi*\psi$ | true | starting state $s_0$ | $\varphi$ true if it is false <br> $\psi$ true if it is false |
| 7 | $\varphi=>\psi$ | false | starting state $s_0$ | $\varphi$ opposite to the actual result <br> $\psi$ opposite to the actual result |
| 8 | $\varphi=>\psi$ | true | starting state $s_0$ | $\varphi$ false <br> $\psi$ true |
| 9 | $\varphi<=>\psi$ | false | starting state $s_0$ | $\varphi$ opposite to the actual result <br> $\psi$ opposite to the actual result |
| 10 | $\varphi<=>\psi$ | true | starting state $s_0$ | $\varphi$ opposite to the actual result <br> $\psi$ opposite to the actual result |
| 11 | $\boldsymbol{AX}\,\varphi$ | false | any successor of $s_0$ holding $\varphi$ | false |
| 12 | $\boldsymbol{AX}\,\varphi$ | true | any successor of $s_0$ not holding $\varphi$ | true |
| 13 | $\boldsymbol{AX_a}\,\varphi$ | false | any successor of $s_0$ in automaton $\boldsymbol{a}$, holding $\varphi$ | false |
| 14 | $\boldsymbol{AX_a}\,\varphi$ | true | any successor of $s_0$ in automaton $\boldsymbol{a}$, not holding $\varphi$ | true |
| 15 | $\boldsymbol{AF}\,\varphi$ | false | any state holding $\varphi$ | false |
| 16 | $\boldsymbol{AF}\,\varphi$ | true | any member of strongly connected subgraph in which $\varphi$ is not satisfied | true |
| 17 | $\boldsymbol{AG}\,\varphi$ | false | starting state $s_0$ | false |
| 18 | $\boldsymbol{AG}\,\varphi$ | true | any state not holding $\varphi$ | true |
| 19 | $\boldsymbol{A}(\varphi\boldsymbol{U_w}\,\psi)$ | false | any state holding $\varphi$ and $\psi$ or a state holding $\varphi$, last in a cycle (before $s_0$) | $\varphi$ false <br> $\psi$ false if it is true |
| 20 | $\boldsymbol{A}(\varphi\boldsymbol{U_w}\,\psi)$ | true | any state holding neither $\varphi$ nor $\psi$ | $\varphi$ true <br> $\psi$ true |
| 21 | $s_0\colon\varphi$ | false | the state $s_0$ | false |
| 22 | $s_0\colon\varphi$ | true | the state $s_0$ | true |
| 23 | $\forall s\in S;\ \varphi$ | false | any state of $S$ satisfying $\varphi$ | false |
| 24 | $\forall s\in S;\ \varphi$ | true | any state of $S$ | true |

| | | | not satisfying $\varphi$ | |
|---|---|---|---|---|
| 25 | $\exists s \in S; \varphi$ | false | any state of $S$ satisfying $\varphi$ | false |
| 26 | $\exists s \in S; \varphi$ | true | any state of $S$ not satisfying $\varphi$ | true |

The program TempoRG [Dasz02] that evaluates temporal formulas, constructs also the critical tree. To do this, the evaluation is performed for the second time with the optimizations disabled:

- The sequence must begin in the specific state: the state finishing the sequence for embracing operator. Therefore the evaluation must not be performed for many states simultaneously.
- The sequence must end in a state appointed by the rules for the given operator (see the table). But the optimization that prevents evaluation of a subformula for a given states many times collecing the results of evaluation for future usage. For example, during the evaluation of $s_0$: **AF** $\varphi$ it may be stored in the data structures that the formula **AF** $\varphi$ is false in the whole future of the state $s_0$ and the evaluation terminates. But the rules of building the sequence say that a state belonging to a strongly connected subgraph must be found, and $s_0$ may not be the member. Therefore, this oprimization should be turned off.
- The critical tree must explain why the subformula gives the erroneous result. In normal evaluation, if the factor is false then the other factor of the conjunction need not be evaluated. But during the construction of the critical tree it must be checked if both factors give false result, and if so, why. This leads to the conclusion that lazy evaluation should not be applied.

Having the starting state of the sequence $s$ specified and the ending state $s'$ found, the sequence of states between them should be calculated. It is done in two steps:

1. Using CBS rule, the sequence of spheres[2] [Dasz01] is constructed from s:
   - $SRC=\{s\} \cup FUT(s)$
   - $cond1=s'$ in sphere; $cond1res$ does not care
   - $cond2=true; cond2res$ does not care
2. Having found the sphere $SPH_n(s)$; the backtracking is performed: any predecessor in the sphere $SPH_{i-1}(s)$ of the chosen state in the sphere $SPH_i(s)$ is teken. For the sphere $SPH_n(s)$ the state $s'$ is chosen. The sequence of chosen states in spheres from $SPH_0(s)$ to $SPH_n(s)$ is just the searched sequence from $s$ to $s'$.

## 4. EXAMPLE 9 – THE SIG PROTOCOL

The tests of COSMA environment [Cwww] are performed on the model of SIG [Grab99] protocol used in the plant monitoring system. The model consists of 25 automata and it contains over 14 000 states and over 32 000 arcs. The following formula was evaluated (general path quantifiers **A** are skipped as in the input of TempoRG program [Dasz02], because the quantifier **E** is not used in QsCTL, see Section **Błąd! Nie można odnaleźć źródła odwołania.**):

---

[2] A sphere is a set of states reachable by given number of arcs from $s$. Sphere $SPH_0(s)$ consists of the state $s$ itself. Sphere $SPH_1(s)$ contains successors of state $s$ (other than $s$). Sphere $SPH_i(s)$ contains these successors of states belonging to $SPH_{i-1}(s)$ which do not belong to 'previous' spheres $SPH_j(s)$, $j<i$. Thus, any sphere $SPH_i(s)$ contains states reachable from state $s$ by $i$ arcs, excluding states belonging to spheres $SPH_j(s)$, $j<i$.

$\forall s \in \{\textbf{\textit{SocketSocket}}.notConnected\};$
  $(\textbf{\textit{X}} \neg CGVar)\ \textbf{\textit{U}}_\textbf{w}$
  $(((\neg CStartVal) * (\neg (\textbf{\textit{in SocketSocket}}.notConnected)))\ \textbf{\textit{U}}_\textbf{w}\ SetVarsOkFlg)$  ∎

**Note.** The following notation is used for operators in the present report and in the TempoRG interface:

| Operator | Report | TempoRG |
|---|---|---|
| Boolean negation | $\neg$ | ! |
| conjunction | * | * |
| disjunction | + | + |
| state $p$ in automaton $\textbf{\textit{a}}$ | $\textbf{\textit{a}}$.p | a.p |
| staying in state $s$ | $\textbf{in}$ s | in s |
| signal $\underline{x}$ being generated | $x$ | x |
| next state (modality) | $\textbf{X}$ | N |
| weak until (modality) | $\textbf{\textit{U}}_\textbf{w}$ | U |
| general path quantifier | *(implied)* | *(implied)* |
| general state quantifier | $\forall$ | A |
| state variable $s$ | s | s |

The meaning of the formula is irrelevant; the only important feature is that the formula is quite complicated and contains several two-argument operators ($\textbf{\textit{U}}_\textbf{w}$, *, $\textbf{\textit{U}}_\textbf{w}$)

The parsing tree of the formula is presented in Fig. 1. Nodes of the tree are numbered following the left order. From now on, every feature raletd to given subformula will be marked with its number.

The result of evaluation is false, and according to the rules presented in the table, a critical tree is constructed. Sequences for nodes ❽ and ❾ are not constructed since the values of these subformulas are just as desired or do not influence the result.

As the sequences must "stick" in the tree, the tree may be presented as in Fig. 2. The sequences are represented as edges of the tree. A node of the tree corresponds to the state that finishes the preceding sequence shown above it (except for the root) and starts the sequences below it (except for the leaves).

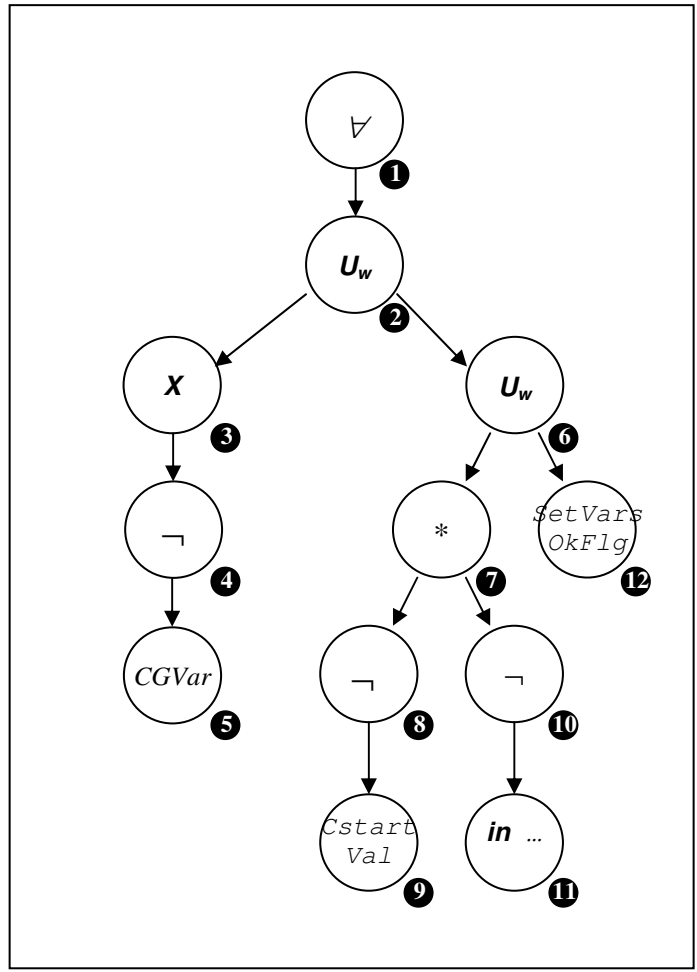

**Fig. 1. Example parsing tree**

The following rules were used during the construction of the critical tree (numbers refer to the table of rules):

| Node | Rule | Desired result of left argument | Desired result of right argument |
|---|---|---|---|
| ❶ $\not\vdash$ | 24 | | ❷ true |
| ❷ $U_w$ | 20 | ❸ true | ❻ true |
| ❸ $X$ | 12 | | ❹ true |
| ❹ $\neg$ | 2 | | ❺ false |
| ❺ *CGVar* | | | |
| ❻ $U_w$ | 20 | ❼ true | ⓬ true |
| ❼ $*$ | 6 | ❽ true (and is) | ❿ true |
| ❽ $\neg$ | not constructed (value as desired) | | |
| ❾ *CstartVal* | not constructed (not reached by the algorithm) | | |
| ❿ $\neg$ | 2 | | ⓫ false |
| ⓫ *in **SocketSocket**.NotConnected* | | | |
| ⓬ *SetVarsOkFlg* | | | |

Because there are many single-state sequences consisting of the same state in the tree (for example four sequences consists of the state $s_4$), the tree may be compressed. Fig. 3 shows the compressed tree. Multi-state sequences (for temporal operators) are represented as edges of the tree. Single-state sequences are contained in nodes.

The critical tree generated by the TempoRG program has one of four possible views:
- states,
- states in automata,
- signals,
- XML [Bray98].

The picture of the views is presented below. In the first three views, the program outputs the parsing tree in an indent form (every level of the tree increments indentation). A sequence for an operator is listed below it. All states in the sequence are marked "OK" except the last one marked "ERROR".

**"States" view** (Fig. 4) shows states of the state space as the elements of the sequence.

**"States in automata" view** (Fig. 5) shows columns representing individual automata. Names of automata are shown in the header. In every column, the state in the automaton is shown only if it changes in comparison to the previous global state.

**"Signals" view** (Fig. 6) is similar to states in automata view, but names of local states are replaced by names of signals generated in states. Names of automata "listening" to the signal follow the name of signal. The automaton listens to the signal if the signal is present in a formula on a label of any arc outgoing from the current state. As before, the header contains the names of automata.

The last output form is **"XML" view** (Fig. 7). It is the synthesis of the three previous forms: global states, states in automata and signals generated and observed are shown. The output is organized in XML format. Fig. 7 contains the structure of the XML view and the fragment of example critical tree.

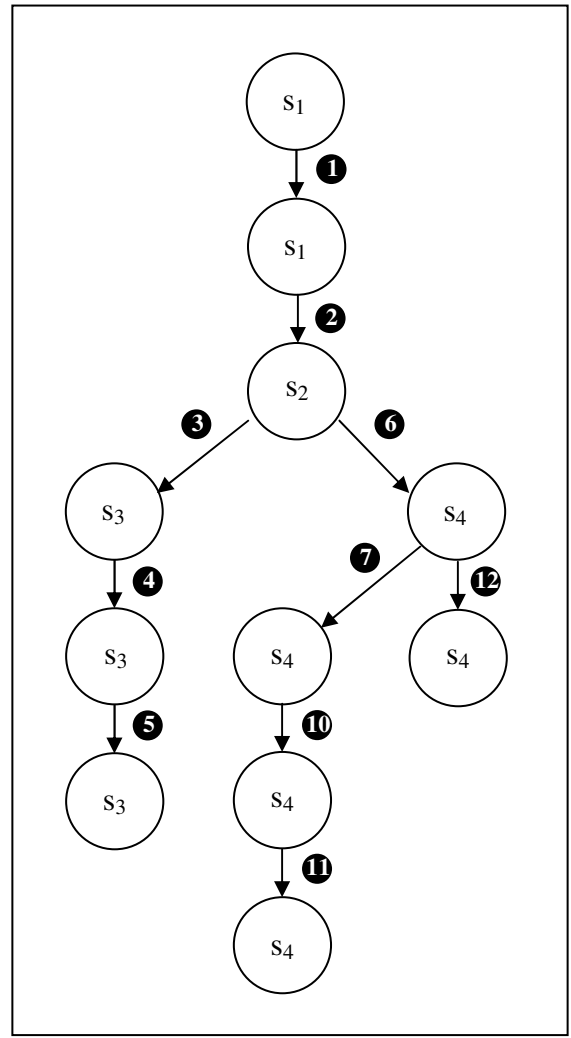

**Fig. 2. Critical tree**

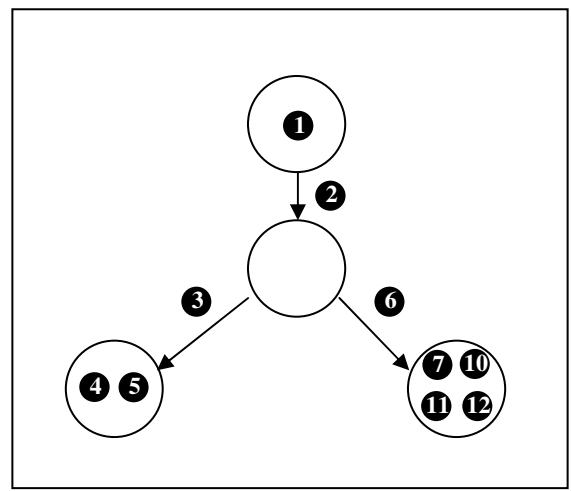

**Fig. 3. Compressed critical tree**

In the TempoRG program, a critical tree is calculated automatically after the negative result of evaluation is achieved. The designer decides on the view of the tree. Basing on the contents of the critical tree, the designer may follow the sequence and identify the reason of the error found. It is the most important practical support for the designer, which allows her/him to redesign the system fixing the incorrect behavior.

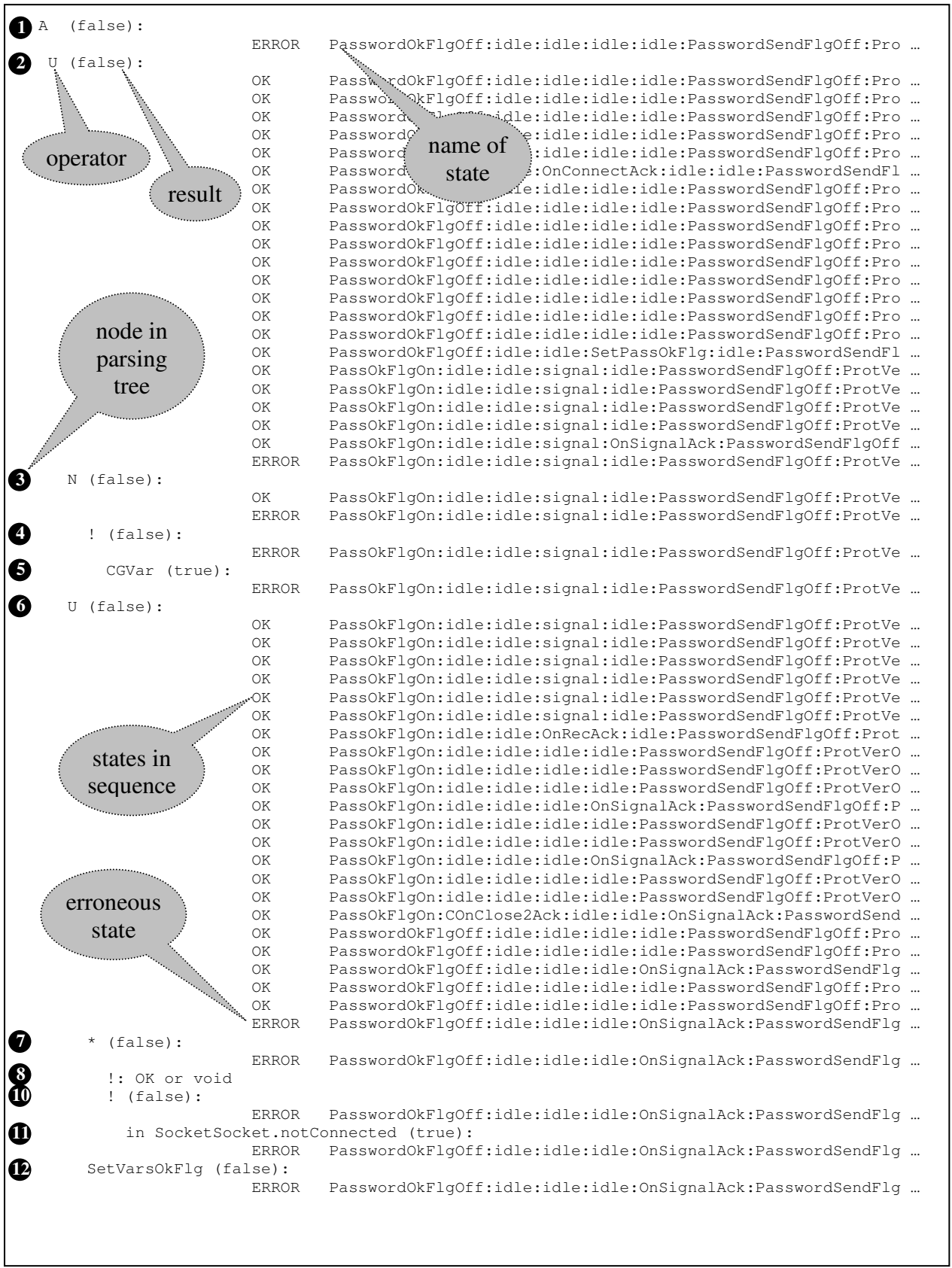

**Fig. 4. „States" view**

**Fig. 5. „States in automata" view**

```
                        PasswordOkFlag              |PasswordOnClose        ...
❶ A  (false):
                        ERROR                       |                       ...
❷ U  (false):
                        OK                          |                       ...
                        OK                          |                       ...
                        OK                          |                       ...
                        OK                          |                       ...
                        OK                          |                       ...
                        OK                          |                       ...
                        OK                          |                       ...
                        OK                          |                       ...
                        OK                          |                       ...
                        OK                          |                       ...
                        OK                          |                       ...
                        OK                          |                       ...
                        OK                          |                       ...
                        OK                          |                       ...
                        OK    PassOkFlg ->VariablesOnReceiv|                ...
                        OK    PassOkFlg ->VariablesOnReceiv|                ...
                        OK    PassOkFlg ->VariablesOnReceiv|                ...
                        OK    PassOkFlg ->VariablesOnReceiv|                ...
                        OK    PassOkFlg ->VariablesOnReceiv|                ...
                        ERROR PassOkFlg ->VariablesOnReceiv|                ...
❸    N  (false):
                        OK    PassOkFlg ->VariablesOnReceiv|                ...
                        ERROR PassOkFlg ->VariablesOnReceiv|                ...
❹    !  (false):
                        ERROR PassOkFlg ->VariablesOnReceiv|                ...
❺      CGVar (true):
                        ERROR PassOkFlg ->VariablesOnReceiv|                ...
❻    U  (false):
                        OK    PassOkFlg ->VariablesOnReceiv|                ...
                        OK    PassOkFlg ->VariablesOnReceiv|                ...
                        OK    PassOkFlg ->VariablesOnReceiv|                ...
                        OK    PassOkFlg ->VariablesOnReceiv|                ...
                        OK    PassOkFlg ->VariablesOnReceiv|                ...
                        OK    PassOkFlg ->VariablesOnReceiv|                ...
                        OK    PassOkFlg ->VariablesOnReceiv|                ...
                        OK    PassOkFlg ->VariablesOnSignal|                ...
                        OK    PassOkFlg ->VariablesOnReceiv|                ...
                        OK    PassOkFlg ->VariablesOnReceiv|                ...
                        OK    PassOkFlg ->VariablesOnReceiv|                ...
                        OK    PassOkFlg ->VariablesOnReceiv|                ...
                        OK    PassOkFlg ->VariablesOnReceiv|                ...
                        OK    PassOkFlg ->VariablesOnReceiv|                ...
                        OK    PassOkFlg ->VariablesOnReceiv|COnClose2Ack ->SocketSocket, Cl ...
                        OK                          |                       ...
                        OK                          |                       ...
                        OK                          |                       ...
                        OK                          |                       ...
                        OK                          |                       ...
                        ERROR                       |                       ...
❼    *  (false):
                        ERROR                       |                       ...
❽      !: OK or void
❿      !  (false):
                        ERROR                       |                       ...
⓫       in SocketSocket.notConnected (true):
                        ERROR                       |                       ...
⓬ SetVarsOkFlg (false):
                        ERROR                       |                       ...
```

Callouts: name of automaton · operator · result · name of signal · addressee · node in parsing tree · states in sequence · separator · erroneous state · no signal generated

**Fig. 6. „Signals" view**

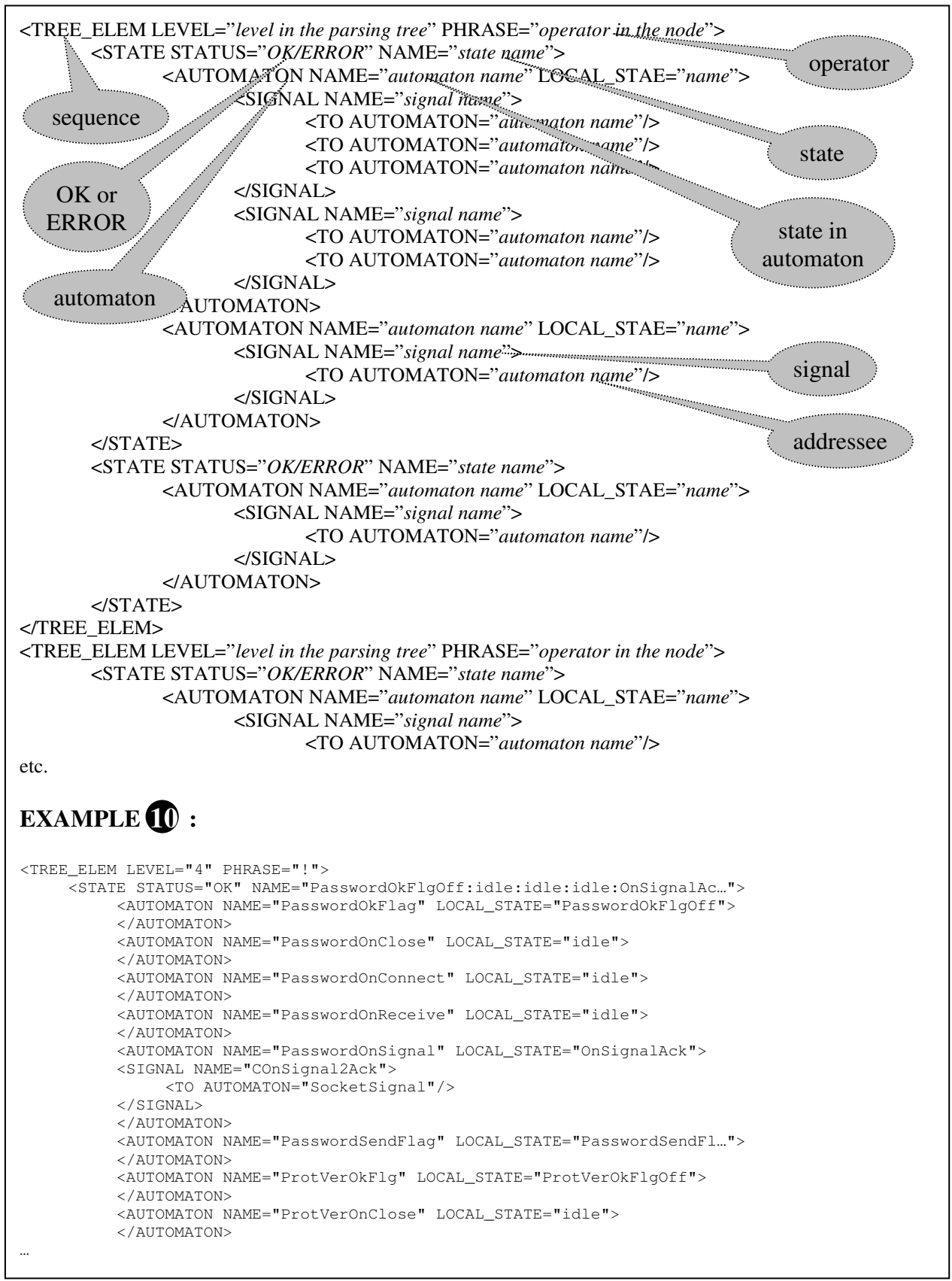

```
<TREE_ELEM LEVEL="level in the parsing tree" PHRASE="operator in the node">
        <STATE STATUS="OK/ERROR" NAME="state name">
                <AUTOMATON NAME="automaton name" LOCAL_STAE="name">
                        <SIGNAL NAME="signal name">
                                <TO AUTOMATON="automaton name"/>
                                <TO AUTOMATON="automaton name"/>
                                <TO AUTOMATON="automaton name"/>
                        </SIGNAL>
                        <SIGNAL NAME="signal name">
                                <TO AUTOMATON="automaton name"/>
                                <TO AUTOMATON="automaton name"/>
                        </SIGNAL>
                </AUTOMATON>
                <AUTOMATON NAME="automaton name" LOCAL_STAE="name">
                        <SIGNAL NAME="signal name">
                                <TO AUTOMATON="automaton name"/>
                        </SIGNAL>
                </AUTOMATON>
        </STATE>
        <STATE STATUS="OK/ERROR" NAME="state name">
                <AUTOMATON NAME="automaton name" LOCAL_STAE="name">
                        <SIGNAL NAME="signal name">
                                <TO AUTOMATON="automaton name"/>
                        </SIGNAL>
                </AUTOMATON>
        </STATE>
</TREE_ELEM>
<TREE_ELEM LEVEL="level in the parsing tree" PHRASE="operator in the node">
        <STATE STATUS="OK/ERROR" NAME="state name">
                <AUTOMATON NAME="automaton name" LOCAL_STAE="name">
                        <SIGNAL NAME="signal name">
                                <TO AUTOMATON="automaton name"/>
```

etc.

## EXAMPLE ❿ :

```
<TREE_ELEM LEVEL="4" PHRASE="!">
        <STATE STATUS="OK" NAME="PasswordOkFlgOff:idle:idle:idle:OnSignalAc…">
                <AUTOMATON NAME="PasswordOkFlag" LOCAL_STATE="PasswordOkFlgOff">
                </AUTOMATON>
                <AUTOMATON NAME="PasswordOnClose" LOCAL_STATE="idle">
                </AUTOMATON>
                <AUTOMATON NAME="PasswordOnConnect" LOCAL_STATE="idle">
                </AUTOMATON>
                <AUTOMATON NAME="PasswordOnReceive" LOCAL_STATE="idle">
                </AUTOMATON>
                <AUTOMATON NAME="PasswordOnSignal" LOCAL_STATE="OnSignalAck">
                <SIGNAL NAME="COnSignal2Ack">
                        <TO AUTOMATON="SocketSignal"/>
                </SIGNAL>
                </AUTOMATON>
                <AUTOMATON NAME="PasswordSendFlag" LOCAL_STATE="PasswordSendFl…">
                </AUTOMATON>
                <AUTOMATON NAME="ProtVerOkFlg" LOCAL_STATE="ProtVerOkFlgOff">
                </AUTOMATON>
                <AUTOMATON NAME="ProtVerOnClose" LOCAL_STATE="idle">
                </AUTOMATON>
…
```

**Fig. 7. „XML" view**

# 5. SEQUENCE DIAGRAMS

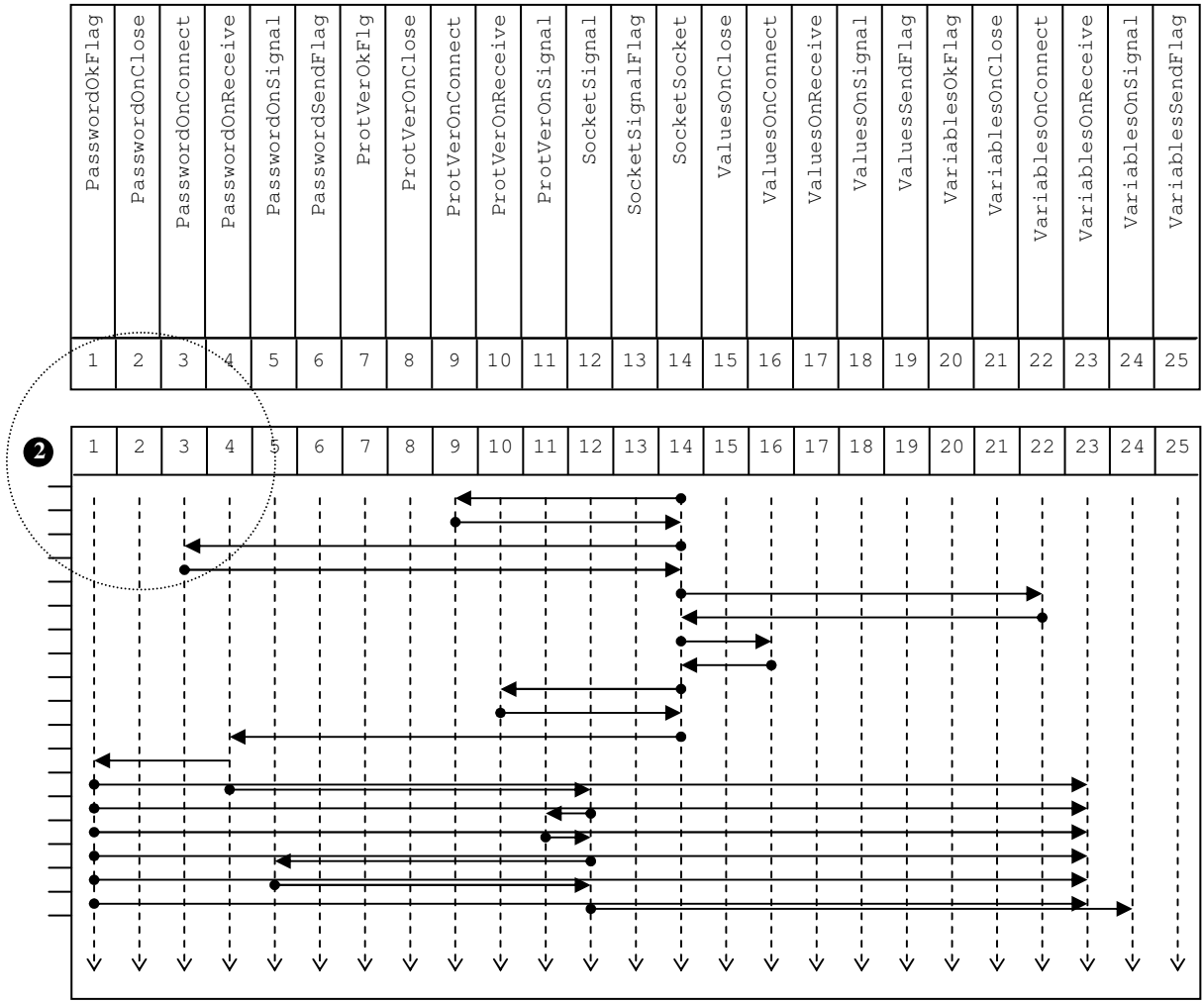

**Fig. 8. Sequence diagram ❷**

To facilitate the further analysis of evaluation results, the XML form may be easily processed by external programs that make further analysis of results of evaluation. For example, UML [OMG99] sequence diagrams may be easily constructed directly from XML output. The sequences for subformulas ❷ and ❻ are presented in the form of sequence diagrams in Figs. 8 and 10 (Fig. 8 contains also a table translating names of automata to codes for simplifications of the next figures; Fig 9 contains the explanation of the elements of a sequence diagram contained in the circle on the left of Fig. 8). Names of signals are not shown for readability. In the sequences, many signals are generated in the same state, so it may be difficult to observe which signals belong to the specific state. To simplify this task, short segments on the left separate distinct states.

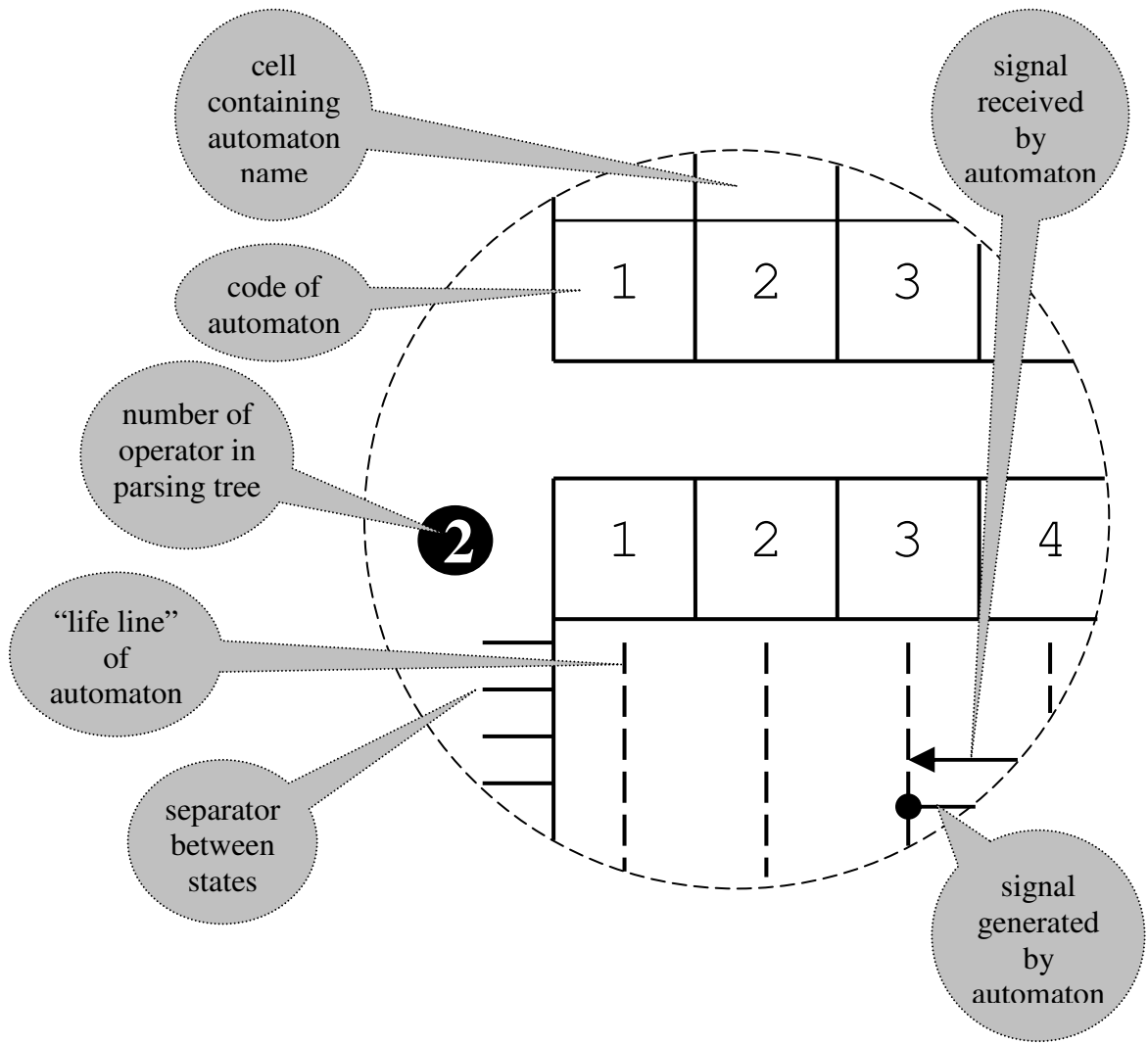

cell containing automaton name

signal received by automaton

code of automaton

number of operator in parsing tree

"life line" of automaton

separator between states

signal generated by automaton

**Fig. 9 Explanation of elements of sequence diagrams (on the example of Fig. 8)**

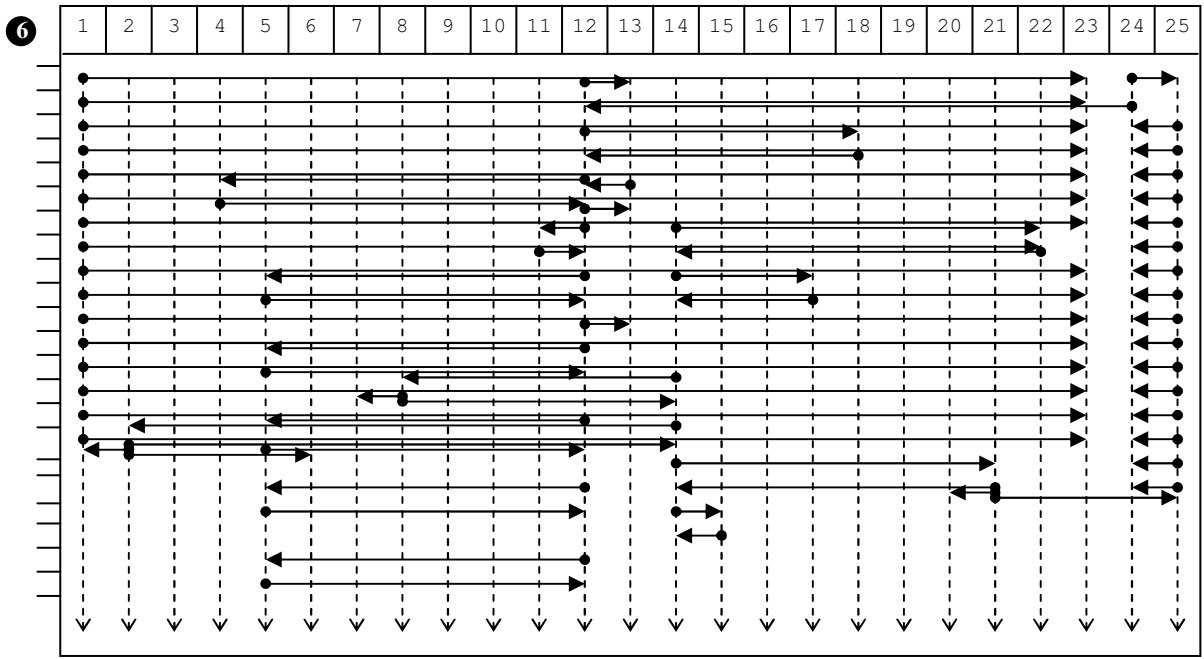

**Fig. 10. Sequence diagram ❻**

In the sequences in which signals are generated continuously state by state, they may be presented as wide arrows like in Fig. 11.

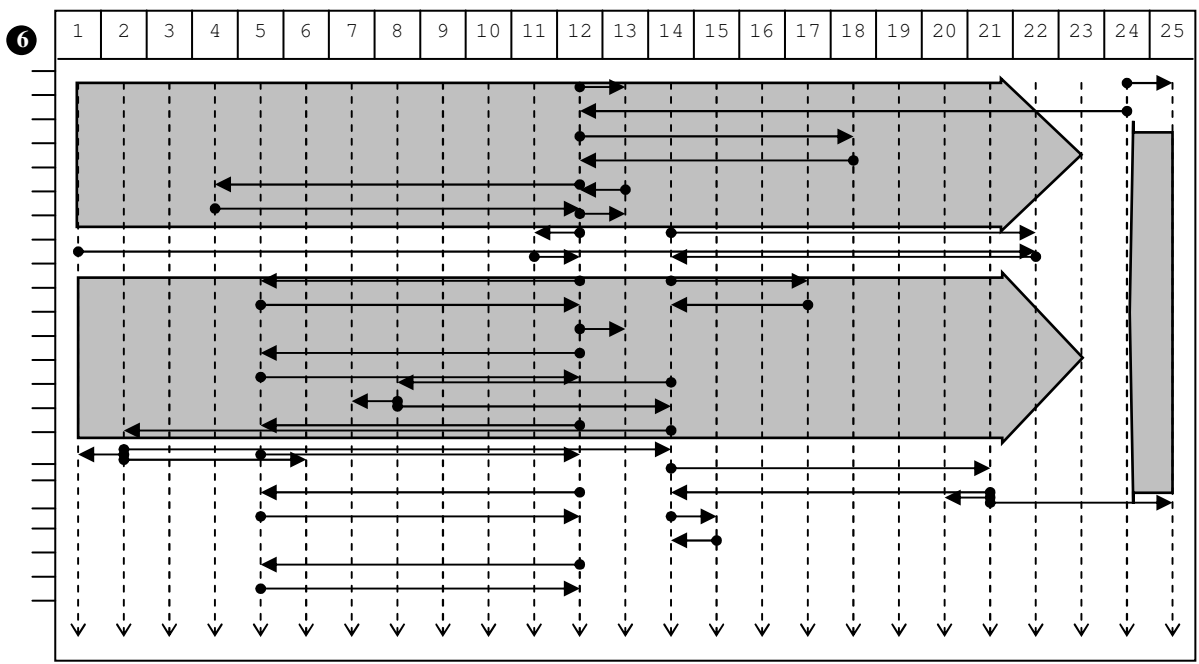

**Fig. 11. Other presentation of sequence diagram ❻**

# 6. CRITICAL TREE VERSUS STATE SPACE REDUCTION

State space reduction (Chapter **Błąd! Nie można odnaleźć źródła odwołania.**) should not be used in combination with searching for a critical tree. The reduction of state space skips the states which give the same value of atomic Boolean formulas and the sequences of states in which given signals are not generated – are compressed to a single state. This makes the critical tree rather useless. To illustrate this observation, the following formula was evaluated over reduced state space:

$\forall s \in \{$ **SocketSocket**.*notConnected* $\}$;
    $(\neg CGVar)$ $\boldsymbol{U_w}$
    $(((\neg CStartVal) * (\neg (in\ \textbf{SocketSocket}.notConnected)))$ $\boldsymbol{U_w}$ *SetVarsOkFlg*) ■

Numbers of nodes in the parsing tree are preserved from the previous example (the node ❸ does not exist in the present example). A sequence in critical tree may be presented as a sequence diagram (like the ones in Figs. 8 and 10), where the signals passed between component automata clearly explain the reason of improper system behavior. For example, a sequence of signals obtained from the state sequence ❷ is listed below.

In the sequence, states are numbered from *1*. If no signal is passed between automata in the state, only the number of the state is present (as in the case of states *1*, *3*, *11*, …). If many signals are passed, they are listed one under another (as in the case of states *17*, *18*, *19*, …). Every signal passed between automata is presented as:

```
source_automaton --{ name_of_signal }--> destination_automaton
```

The sequence is as follows:

```
 1.
 2. SocketSocket --{ COnConnect1 }--> ProtVerOnConnect
 3.
 4. ProtVerOnConnect --{ COnConnect1Ack }--> SocketSocket
 5. SocketSocket --{ COnConnect2 }--> PasswordOnConnect
 6. PasswordOnConnect --{ COnConnect2Ack }--> SocketSocket
 7. SocketSocket --{ COnConnect3 }--> VariablesOnConnect
 8. VariablesOnConnect --{ COnConnect3Ack }--> SocketSocket
 9. SocketSocket --{ COnConnect4 }--> ValuesOnConnect
10. ValuesOnConnect --{ COnConnect4Ack }--> SocketSocket
11.
12.
13. SocketSocket --{ COnRec1 }--> ProtVerOnReceive
14. ProtVerOnReceive --{ COnRec1Ack }--> SocketSocket
15. SocketSocket --{ COnRec2 }--> PasswordOnReceive
16. PasswordOnReceive --{ SetPassOkFlg }--> PasswordOkFlag
17. PasswordOkFlag --{ PassOkFlg }--> VariablesOnReceive
    PasswordOnReceive --{ Signal2 }--> SocketSignal
18. PasswordOkFlag --{ PassOkFlg }--> VariablesOnReceive
    SocketSignal --{ COnSignal1 }--> ProtVerOnSignal
19. PasswordOkFlag --{ PassOkFlg }--> VariablesOnReceive
    ProtVerOnSignal --{ COnSignal1Ack }--> SocketSignal
20. PasswordOkFlag --{ PassOkFlg }--> VariablesOnReceive
    SocketSignal --{ COnSignal2 }--> PasswordOnSignal
21. PasswordOkFlag --{ PassOkFlg }--> VariablesOnReceive
    PasswordOnSignal --{ COnSignal2Ack }--> SocketSignal
22. PasswordOkFlag --{ PassOkFlg }--> VariablesOnReceive
    SocketSignal --{ COnSignal3 }--> VariablesOnSignal
23. PasswordOkFlag --{ PassOkFlg }--> VariablesOnReceive
    SocketSignal --{ SetSigFlg }--> SocketSignalFlag
    VariablesOnSignal --{ SetVarsSendFlg }--> VariablesSendFlag
```

From the sequence one can easily observe how the automaton **SocketSocket** operates groups of other automata responsible for given phases of establishing the connection (**ProtVer**, **Password**, **Variables**, **Values**) and how the automaton **SocketSignal** takes over this role from state *18*. However, the reduction of state space skips the states which give the same value of atomic Boolean formulas and the sequences of states in which given signals are not generated are compressed to a single state. In our example, the sequence ❷ in reduced state space is presented in Fig. 12.

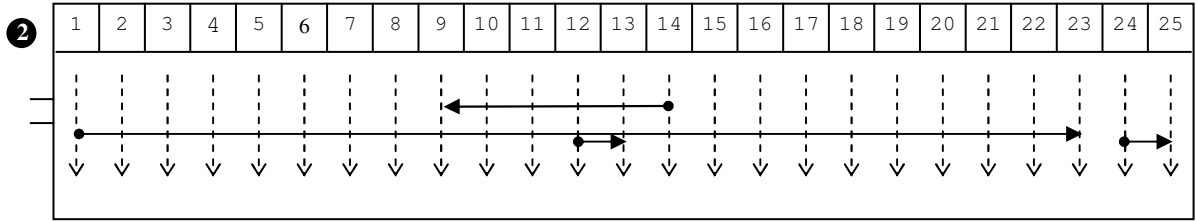

**Fig. 12. Sequence diagram ❷ in reduced state space**

The signals passed between automata in the sequence are:

```
 1.
 2.  SocketSocket --{ COnConnect1 }--> ProtVerOnConnect
23.  PasswordOkFlag --{ PassOkFlg }--> VariablesOnReceive
     SocketSignal --{ SetSigFlg }--> SocketSignalFlag
     VariablesOnSignal --{ SetVarsSendFlg }--> VariablesSendFlag
```

Three states are preserved in reduced state spave out of original 23 states, because:

- the subformula **in SocketSocket**.*notConnected* is true in state *1*, and false in states *2-23* of the original sequence;
- the subformula *CGVar* is true in state *23* and false in states *1-22*;
- all other atomic formulas do not change value along the original sequence.

The new sequence says nothing interesting about the actions that lead to the erroneous state at the end of the path. Therefore, after negative evaluation of the formula over reduced state space, it should be evaluated once again over original state space (and without optimizations) to find a critical tree.

## 7. CONCLUSIONS

In the report, a concept of construction of critical trees for negatively evaluated temporal formulas is presented. Generation of couterexapmples decides of the strength of model checking methods. The designer is informed not only on the existence of fault, but she/he is also precisely informed of the sequence of actions that leads to the erroneous state. The rules for critical tree generation are defined in a way that gives the designer maximum information about the fault. Especially the division of counterexample into sequences responsible for individual operators is useful in redesign of the verified system.

Based on the critical tree, a sequence diagram [OMG99] of error prone message exchange may be visualized, or error simulation over a graphical state space may be executed. The TempoRG program outputs the counterexample in a form that may be easily converted to the graphical form of sequence diagram or simulated dynamically as state changes in a fragment of RG. Such a result analyzer is planned in the COSMA environment.

## References

[Amma98] Ammann P.E., , Black P. E., Majurski W.,1998, "Using Model Checking to Generate Tests from Specifications", *Proc. 2$^{nd}$ IEEE Int. Conf. on Formal Engineering Methods (ICFEM'98)*, Brisbane, Australia (Dec 1998), Staples J., Hinchey M. G., and Liu S. (eds.), IEEE Computer Society, pp. 46-54.

[Bray98] Bray T., Paoli J., Sperberg-McQueen C, 1998, Extensible markup language (XML) 1.0, http://www.w3c.org/TR/REC-xml, W3C Recommendation

[Chan98] Chan W., Anderson R. J., Beame P., Burns S., Modugno F., Notkin D., Reese D., 1998, "Model checking large software specifixations", *IEEE Transactions on Software Engineering*, SE-24(7), Jul 1998, pp. 498-520

[Clar86] Clarke E. M., Emerson E. A., Sistla A. P., 1986, "Automatic Verification of Finite State Concurrent Systems Using Temporal Logic Specifications", *ACM Transactions on Programming Languages and Systems*, 8(2) (April 1986), pp. 244-263

[Clar89] Clarke E. M., Grumberg O., Kurshan R. P., 1989, "A Synthesis of Two Approaches for Verifying Finite State Concurrent Systems", in *Proc. Of Symposium on Logical Foundations of Computer Science: Logic at Botik '89*, *Lecture Notes in Computer Science* vol. 363, Springer-Verlag, New York

[Clar99] Clarke E. M., Grumberg O., Peled D., 1999, *Model Checking*, MIT Press 1999

[Cwww] http://www.ii.pw.edu.pl/cosma

[Dasz00] Daszczuk W. B., 2000, "State Space Reduction For Reachability Graph of CSM Automata", *Institute of Computer Science, WUT, ICS Research Report* No 10/2000

[Dasz01] Daszczuk W. B., 2001, "Evaluation of Temporal Formulas Based on Checking By Spheres", *Proc. Euromicro Symposium on Digital Systems Design – Architectures, Methods and Tools*, September 4-6, Warsaw, Poland, pp. 158-164

[Dasz02] Daszczuk W. B., 2002, „*Verification of Temporal Properties in Concurrent Systems*", PhD Thesis, Warsaw University of Technology

[Grab99] Grabski W., Daszczuk W. B., Mieścicki J., Dobrowolski H., 1999, „Verication of Event Protocol of Establishing and Closing of a Connection in ESS System", *Institute of Computer Science, WUT, ICS Research Report* No 10/99

[Holz97] Holzman G. J., 1997, The model checker SPIN, *IEEE Transactions on Software Engineering*, SE-23(5), May 1997, pp. 279-295

[Jans98] Janssen G. L. J. M., 1998, "*Logics for Digital Circuit Verification: Theory, Algorithms and Applications*", PhD thesis, Eindhoven University of Technology, 1998

[Kupf98] Kupferman O., Vardi M. Y., 1998, "Weak Alternating Automata and Tree Automata Emptiness", in *Proc. of 30$^{th}$ Annual ACM Symposium on Theory of Computing*, Dallas, Texas, USA, May 23-26, 1998. ACM, pp.224-233

[McMi92] McMillan K. L., 1992, "*Symbolic Model Checking, An Approach to the State Explosion Problem*". PhD thesis, School of Computer Science, Carnegie Mellon University, Pittsburgh, PA, 1992

[OMG99] OMG, *Unified Modeling Language Specification*, Version 1.3, June 1999

[Pasa01] Păsăreanu C.S., Dwyer M. B., Visser W., 2001, "Finding Feasible Counter-examples whan Model Checking Abstracted Java Programs", *Proc. 7th Int.*

Conf. Tools and Algorithms for the Construction and Analysis of Systems TACAS 2001, Genova, Italy, April 2-6, 2001, LNCS 2031, p. 284

[Vard96]   Vardi M. Y., "An Automata-Theoretic Approach to Linear Temporal Logic", in *Logics for Concurrency*, Lecture Notes in Computer Science vol. 1043, Springer-Verlag, New York, 1996, pp. 238-266

[Vard98]   Vardi M. Y., "Sometimes and Not Never Re-revisited: On Branching Versus Linear Time", in *Proc. of 9th International Conference on Concurrency Theory*, CONCUR 1998, Nice, France, September 8-11, 1998, Lecture Notes in Computer Science vol. 1466, Springer-Verlag, New York, 1998, pp. 1-17,

[Viss97]   Visser W., Barringer H., Fellows D., Gough G., Williams A., "Efficient CTL* Model Checking for Analysis of Rainbow Designs", in *Proc. of CHARME '97*, Montreal, October 1997, pp. 128-145

[Viss00]   Visser W., Barringer H., "Practical CTL* Model Checking: Should SPIN be extended?", in *International Journal on Software Tools for Technology Transfer* Vol. 2 (2000) No 4, Special Section on SPIN, pp. 350-365